\documentclass[useAMS,usenatbib]{mn2e}

\usepackage{graphicx}
\usepackage{aas_macros}
\usepackage{amsmath}
\def\Mo{M_\odot}

\def\apjs{ApJS}
\def\vrf{$v_{\rm rf}$}
\def\kms{{\rm km\,s^{-1}}}
\begin{document}
\title[Trajectories of hypervelocity stars]{The nature of hypervelocity stars as inferred from their galactic trajectories}
\author[K.~M. Svensson et al.]
{Karl~M.~Svensson\thanks{E-mail: K.M.Svensson@warwick.ac.uk}, Ross~P.~Church,
Melvyn~B.~Davies \\
Lund Observatory, Box 43, SE-221 00, Lund, Sweden.}

\date{Accepted 2007 September 24. Received 2007 September 21; in original form 2007 August 02}

\pagerange{\pageref{firstpage}--\pageref{lastpage}} \pubyear{2007}

\maketitle

\label{firstpage}

\begin{abstract}

We have computed the galactic trajectories of twelve hypervelocity stars (HVSs)
under the assumption that they originated in the Galactic Centre.  We show that
eight of these twelve stars are bound to the Galaxy.  We consider the subsequent
trajectories of the bound stars to compute their characteristic orbital period,
which is 2\,Gyr.  All eight bound stars are moving away from the centre of the
Galaxy, which implies that the stars' lifetimes are less than 2\,Gyr.  We thus
infer that the observed HVSs are massive main sequence stars, rather than blue
horizontal branch stars.  The observations suggest that blue HVSs are ejected
from the Galactic Centre roughly every 15\,Myr.  This is consistent with the
observed population of blue stars in extremely tight orbits round the central
super-massive black hole (SMBH), the so-called S-stars, if we assume that the
HVSs are produced by the breakup of binaries. One of the stars in such a binary
is ejected at high velocities to form a HVS; the other remains bound to the SMBH
as an S-star.

We further show that the one high-velocity system observed to be moving towards
the Galactic Centre, SDSS J172226.55+594155.9, could not have originated in the
Galactic Centre; rather, we identify it as a halo object.

\end{abstract}


\section{Introduction}

%

The existence of stars moving at velocities several times higher than the
escape velocity of the galaxy was first proposed by
\citet{1988Natur.331..687H}.  These {\it hypervelocity} stars (HVSs) would have
been ejected from the very centre of the Galaxy through interactions between
binary stars and the central black
hole. Since the first discovery of an HVS in 2005 several more have been
found. \citet{2007ApJ...660..311B} and \citet{2006ApJ...647..303B} report the
results of a targeted survey for HVSs. Candidate early-type stars are identified
with SDSS photometry and observed with the 6.5 MMT telescope or the Whipple 1.5
m Tillinghast telescope. Precise spectral types and heliocentric radial
velocities are measured from the obtained stellar spectra. The evolutionary
stage is undetermined; stars of the observed colours could be on either the main
sequence or blue horizontal branch.  The survey is complete to a distance of 100
to 150 kpc for B-type stars.  \citet{2007ApJ...660..311B} report that 11 stars
are believed to belong to a population of HVSs ejected from the Galactic Centre
on bound orbits on the basis of measurement of their radial velocities.  The
radial velocities are transformed into the galactic restframe via
\begin{multline}
v_{\rm rf}=v_{\rm r}+(10 \cos l \cos b + 5.2 \sin l \cos b + 7.2 \sin b )\kms \\
 +220\,\kms \sin l \cos b. 
\label{eq:vrf}
\end{multline}
where $v_{\rm r}$ is the heliocentric radial velocity and $l$ and $b$ are the
galactic longitude and latitude respectively.  Thus $v_{\rm rf}$ is the
component of the galactic restframe velocity of a star that lies along the line
connecting the star to the Sun.  As only this component of the velocity is
known, the star's Galactic restframe velocity must be at least $v_{\rm rf}$. 

A plot of $v_{\rm rf}$ versus observed number of stars for a clean sample of
B-type stars can be seen in figure 3 of \citet{2007ApJ...660..311B}.
The majority of stars have \vrf{} in the range $-275\,\kms <v_{\rm
rf}< 275\,\kms$. This is consistent with them being either halo or disc stars.
Eleven stars, however, are found with larger $v_{\rm rf}$, between $275 {\rm
km\,s^{-1}}$ and $450 {\rm km\,s^{-1}}$.  One star is found to have a comparably
large but negative $v_{\rm rf}$, SDSS J172226.55+594155.9, which has $v_{\rm
rf}=-286 {\rm km\,s^{-1}}$.  In our sample of possibly-bound hypervelocity stars
we also include HVS 7, SDSS J113312.12+010824.9 \citep{2006ApJ...647..303B},
since it has $v_{\rm rf}=418$ km/s and the escape velocity at its current
position is $421\,\kms$.

In this paper we consider the origin and nature of these stars. We use their
galactocentric origin to calculate their space velocities and subsequent
trajectories, which allows us to determine whether they are bound to the Galaxy.

\section{Producing hypervelocity stars}


In the ejection mechanism suggested by \citet{1988Natur.331..687H} a
stellar-mass binary approaches the supermassive black hole (SMBH), Sgr A*. The
binary is tidally disrupted by interaction with the SMBH, leading to the capture
of one of its stars and the ejection of the other. Conservation of energy leads
to extreme ejection velocities as the captured star is extremely tightly bound
to the black hole. These captured stars are suggested as the origin of the
so-called S-stars found in the Galactic Centre.

\citet{2006ApJ...653.1194B} have performed numerical simulations of the ejection
mechanism.  They derive the following relation between ejection velocity,
$v_{\rm ej}$,
binary separation, $a_{\rm bin}$, masses of the binary components, $M_1$ and
$M_2$, and the mass of the black hole $M_{\rm bh}$:
\begin{multline}
v_{\rm ej}=1760 \left( \frac{a_{\rm bin}}{0.1\,{\rm AU}} \right)^{-0.5} \left( \frac{M_1+M_2}{2\,
    M_{\odot}}\right)^{1/3}\\
\times  \left( \frac{M_{\rm bh}}{3.5 \times 10^6 M_{\odot}}
\right)^{1/6} f_{\rm r} \,{\rm km\,s^{-1}}.
\label{eq:vej}
\end{multline}
The ejection velocity depends on $f_{\rm r}$, given by
\begin{multline}
f_{\rm r}=0.774+0.0204D-6.23 \times10^{-4}D^2 +7.62 \times10^{-6}D^3\\
-4.24 \times10^{-8}D^4+8.62 \times 10^{-11} D^5,
\label{eq:f_r}
\end{multline}
where $D$ is the Hills parameter, defined as
\begin{equation}
\label{eq:D}
D= \left( \frac{R_{\rm min}}{a_{\rm bin}} \right) \left( \frac{2 M_{\rm bh}}{10^6 (M_1+M_2)} \right)^{-1/3}.
\end{equation}
Here $R_{\rm min}$ is the closest approach between the binary and the black
hole and $a_{\rm bin}$ is the semimajor axis of the binary.
The Hills parameter also relates to the probability of an interaction leading
to an ejection, $P_{\rm ej}$, via
\begin{equation}
\label{eq:P_ej}
P_{\rm ej} \approx 1-\frac{D}{175}.
\end{equation}
The case $R_{\rm min}=0$ corresponds to $D=0$; interesting encounters have $D$
between 0 and 175.  This leads to values of $f_{\rm r}$ in the range $0.5 <
f_{\rm r} < 1$.

\section{Trajectories of hypervelocity stars}

%
If \vrf{} is greater than the galactic escape velocity at the position of the
HVS then it is unbound.  For HVSs with slightly lower observed velocities the
lack of measured proper motions precludes a solid determination of whether they
are bound or unbound.  However, we know that the HVSs have been ejected from the
centre of the Galaxy.  Hence we determine the range of proper motions that give
them trajectories that pass through the Galactic centre.  Thus equipped with
full space velocities we are able to determine if the stars are bound or
unbound.

To determine their trajectories we add components of proper motion to the known
heliocentric radial velocities. The velocity is transformed into galactic
$(U,V,W)$ velocities according to \citet{1987AJ.....93..864J}. The equations of
motion are then integrated in a model of the galactic potential.  We use the
model suggested by \citet{1990ApJ...348..485P}.  This consists of two
Miyamoto-Nagai potential terms which represent the disc and bulge, and a
spherically symmetric halo component.  We truncate the halo density profile to
keep its mass finite.  Following \citet{1999MNRAS.310..645W} we adopt a halo
mass of $1.9 \times 10^{12} M_{\odot}$, which implies truncation of the halo at
a radius of $237\,{\rm kpc}$.  Outside this we model the halo as a central
potential.

We sample proper motions from a grid covering all possible space velocities. The
absolute magnitudes depend on the evolutionary stage, thus the heliocentric
distances will be different depending on whether the stars are on the main
sequence (MS) or blue horizontal branch (BHB). The galactic coordinates $(R,z)$
will also be different, giving different trajectories depending on the
evolutionary stage assumed.  The trajectories are integrated backwards from the
current position to the Galactic Centre, both under the assumption that the
stars are on the MS and that they are on the BHB. 

\section{Results}
\begin{figure}                                               
\centering                                                       
  \includegraphics[width=\columnwidth]{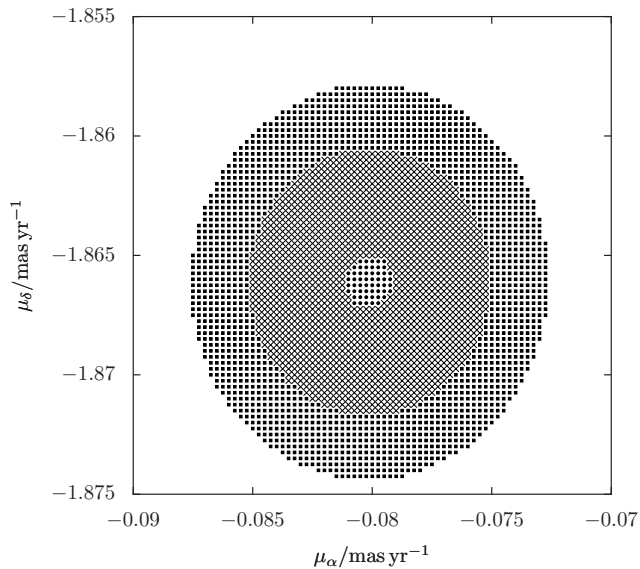}                          
     \caption{Proper motions for SDSS J081828.07+570922.1 that represent
     trajectories that pass close to the Galactic Centre. The outer region of
     solid squares represents proper motions such that the trajectories pass
     within 75\,pc of the Galactic Centre. The dotted region represents proper
     motions with trajectories that pass within 50\,pc of the Galactic Centre
     and the innermost region of circles are proper motions taking the
     trajectory within 10\,pc of the Galactic Centre. For these calculations we
     assume that the star is on the MS.}
\label{fig:pm}                                         
\end{figure}

\begin{figure}                                                  
\centering                                                       
  \includegraphics[width=\columnwidth]{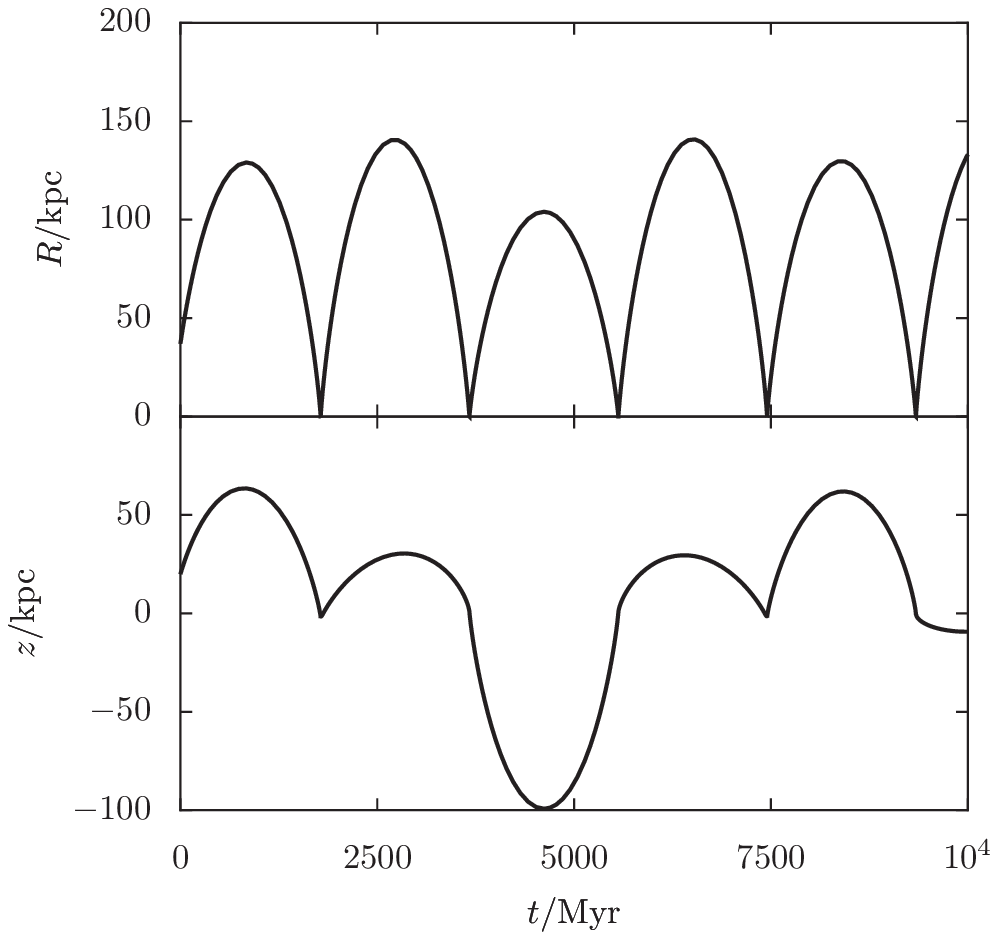}
     \caption{The trajectory of SDSS J081828.07+570922.1 integrated in the
       Galactic potential, assuming that it is a MS star. The cylindrical
       radius, $R$ (top) and height above the plane of the Galaxy, $z$ (lower)
       are plotted against time.  The trajectory starts at the point of ejection
       from the Galactic Centre and continues for the subsequent $10\,{\rm
       Gyr}$.  This trajectory is calculated assuming that the star is on the
       MS.}
\label{fig:Rz}                                         
\end{figure}

%
%
In Figure \ref{fig:pm} we plot proper motions that take a typical star from our
sample, SDSS J081828.07+570922.1, within 75\,pc, 50\,pc and 10\,pc from the
centre of the Galaxy. In Figure \ref{fig:Rz} we plot the trajectory of SDSS
J081828.07+570922.1 that takes it closest to the Galactic Centre.

For each star in our sample of twelve possibly-bound HVSs there exists a small
range of proper motions consistent with ejection from the Galactic Centre;  that
is, for which the corresponding trajectory takes the star through the Galactic
Centre. The twelve HVSs are all found to have been ejected within the last
170\,Myr, with a reasonably uniform distribution in ejection times and a mean
ejection rate of one HVS every 15\,Myr.  \citet{2003ApJ...599.1129Y} predict a
binary breakup rate of $10^{-5}\,{\rm yr}^{-1}$.  Some of these binaries will
contain massive stars, some of which will be ejected and observed as HVSs.
Taking into account selection effects the observed rate is compatible with the
predictions of \citet{2003ApJ...599.1129Y}, as shown by
\citet{2006ApJ...640L..35B}.

The results from our investigation are given in Table~\ref{tab:HVS}. For each
star we consider its position, velocity and trajectory in the cases where it is
on the MS and the BHB.  At the present day we list the proper motions most
consistent with a Galactic Centre origin, the $(R,z)$ coordinates, the lab-frame
speed given these proper motions and the escape speed at the current position.
We also list the ejection speed, which is the speed of the star when its
trajectory crosses the Galactic Centre, and the time since ejection.  We find
that eight out of our twelve stars are bound in the galactic potential.  The
ejection velocities for the bound HVSs are typically about $800\,\kms$.

To estimate the robustness of the results, we let the mass of the galactic halo
 assume other values and then observe which stars in the sample remain unbound.
Reducing the halo mass to half of the best fit value causes only one star, 
SDSS J141723.34+101245.7, to become unbound. Investigating the limits of the 
1-sigma confidence interval of \citet{1999MNRAS.310..645W}, $2.0 \times 10^{11} \Mo$ and
  $5.5 \times 10^{12} \Mo$ respectively, we find that none of the suspected bound
 hypervelocity stars remain bound with halo mass of only  $2.0 \times 10^{11} \Mo$.
 On the other side, a halo mass of $5.5 \times 10^{12} \Mo$ means that
 SDSS J110224.37+025002.8 and SDSS J115245.91-021116.2 become bound.

\begin{table*}
\begin{tabular}{l|r|llllllll}
\hline

ID & &$\mu_{\rm \alpha}$&$\mu_{\rm \delta}$ & $R$ &$z$ &$v$ & $v_{\rm
  esc}$&$v_{\rm ej}$&$t_{\rm ej}$\\

 &&${\rm [mas/yr]}$&${\rm [mas/yr]}$ & ${\rm [kpc]}$ &${\rm [kpc]}$ &${\rm
  [km/s]}$ & ${\rm [km/s]}$&${\rm [km/s]}$&${\rm [Myr]}$\\
\hline
Bound:\\
\hline

SDSS J074950.24+243841.2&
MS&-0.01&-0.55&61.8&23.2&301.&407.&845.&171.\\
& BHB&   -0.34&   -1.71&25.9& 7.8& 305.& 482.& 806.&  69.\\

SDSS J075055.24+472822.9&
MS& 0.06&-1.68&38.3&17.1&312.&446.&830.&106.\\
& BHB&   -1.05&   -6.38&16.3& 4.7& 326.& 518.& 792.&  41.\\

SDSS J075712.93+512938.0&
MS&-0.07&-2.56&29.3&12.8&341.&468.&829.& 76.\\
& BHB&   -3.55&  -12.25&12.7& 2.8& 367.& 540.& 795.&  29.\\

SDSS J081828.07+570922.1&
MS&-0.08&-1.87&37.0&19.9&300.&446.&825.&109.\\
& BHB&   -0.54&   -3.90&22.0& 9.7& 307.& 491.& 802.&  61.\\

SDSS J090710.08+365957.5&
MS&-0.19&-0.87&45.6&34.2&282.&419.&832.&155.\\
& BHB&   -0.73&   -1.97&24.7&15.2& 288.& 476.& 804.&  77.\\

SDSS J140432.38+352258.4&
MS&-1.30&-0.54&14.2&49.0&339.&429.&849.&122.\\
& BHB&   -4.49&   -1.08& 7.5&16.4& 371.& 511.& 816.&  40.\\

SDSS J141723.34+101245.7&
MS&-1.55& 0.07&14.7&45.7&410.&434.&877.&100.\\
& BHB&   -2.88&   -0.55& 1.5&18.9& 340.& 506.& 806.&  45.\\

SDSS J142001.94+124404.8&
MS&-2.34&-0.27& 4.5&26.6&372.&480.&835.& 61.\\
& BHB&   -5.57&   -0.39& 2.5&11.7& 385.& 539.& 804.&  26.\\

\hline
Unbound:\\
\hline 

SDSS J110224.37+025002.8&
MS&-0.81& 0.39&31.9&39.8&432.&429.&890.&102.\\
& BHB&   -2.53&    1.18&16.2&16.4& 461.& 493.& 871.&  44.\\

SDSS J115245.91-021116.2&
MS&-0.98& 0.52&28.8&44.5&445.&425.&898.&104.\\
& BHB&   -3.98&    2.15&11.4&13.9& 494.& 511.& 879.&  32.\\

SDSS J144955.58+310351.4&
MS&-7.44& 0.53& 6.3&14.7&644.&519.&967.& 23.\\
& BHB&  -28.96&   12.88& 6.4& 6.3&1094.& 561.&1292.&   8.\\

SDSS J113312.12+010824.9&
MS&-0.94& 0.71&31.1&46.6&567.&421.&967.& 90.\\
(HVS 7)& BHB&   -5.27&    3.58&11.7&12.3& 646.& 516.& 970.&  24.\\
\hline

\end{tabular}
\caption{Values calculated from our trajectories for the 11 hypervelocity stars
found in the survey of \citet{2007ApJ...660..311B} and HVS 7 (final row).  For
each star the first line (labelled MS) represents the trajectory calculated
assuming that the star is a main-sequence star, and the second line the
trajectory calculated assuming that it is a blue horizontal branch star.  Proper
motions $\mu_\alpha$ and $\mu_\delta$ consistent with the galactic centre
ejection scenario have been found in all cases. The present-day Galactic radius
$R$ and height $z$ are also listed, along with the present-day speed of the star
in the rest frame of the Galaxy, $v$.  The escape velocity at the star's current
position and the velocity with which it was ejected from the Galactic Centre are
$v_{\rm  esc}$ and $v_{\rm  ej}$.  The ejection velocity is taken as the
velocity when the star pass through the Galactic centre.  The time taken to
travel from the Galactic Centre to the present-day position is $t_{\rm ej}$.}
\label{tab:HVS}
\end{table*}

\section{Discussion}

\subsection{The evolutionary stage of bound HVS}

\begin{figure}                                                 
\centering                                                       
  \includegraphics[angle=-90,width=\columnwidth, trim=0mm 30mm 0mm 35mm,clip]{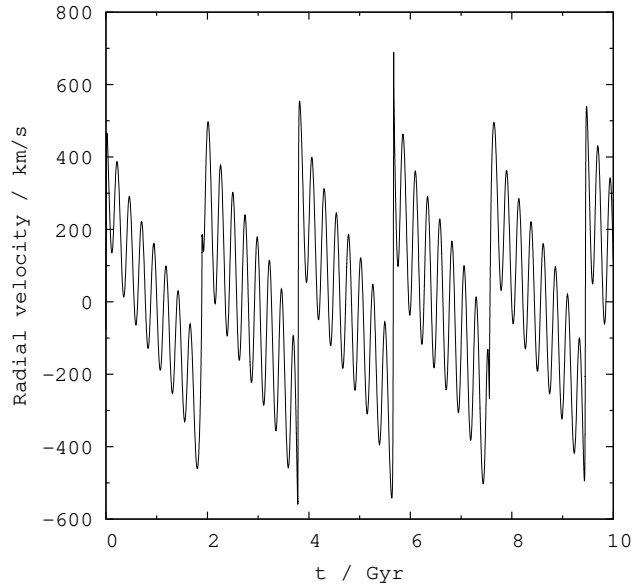}                          
     \caption{The heliocentric radial velocity versus time of SDSS
       J081828.07+570922 from ejection and for the projected trajectory for 10
       Gyrs. The trajectory here is integrated assuming that the star is on the
       MS.}
\label{fig:vr}                                         
\end{figure}

The return time, the time between peri-galacticon passages for a bound HVS, is
typically 2 to 3 Gyr when we assume that the stars are on the MS. If we assume
instead that they are on the BHB the return time is typically 1 to 2 Gyr.
If the stars are on the MS then their colours correspond to masses between 2 and
4 $M_{\odot}$; if they are on the BHB then they can be considerably less
massive.  The MS lifetime is a sensitive function of the stellar mass; low mass
stars are long-lived compared to their massive counterparts. Assuming the stars
are on the MS the least massive, and hence most long-lived, star in Table
\ref{tab:HVS} has a MS lifetime of about 450 Myr. The shortest MS lifetime is
100 Myr.  The orbital periods for these stars are sufficiently long that they
will evolve to become white dwarfs before they return to the Galaxy.  However if
they are BHB stars their progenitors can be of substantially lower mass.  If
they are ejected from the Galactic Centre as low-mass MS stars they can undergo
several orbits before evolving onto the BHB.  In Figure \ref{fig:vr} we plot
heliocentric radial velocities as a function of time for a bound HVS, SDSS
J081828.07+570922.  Over a whole orbit the star spends equal amounts of time
approaching the Galactic Centre as receding from it.  The high-frequency
oscillation has a period of 240\,Myr, and is an effect of the Sun's galactic
orbit.

A low mass HVS, ejected from the Galactic Centre as a MS star, would have an
equal probability of evolving onto the BHB when it is receding from the Galactic
Centre as when it is approaching it.  As no bound HVSs are observed with
negative heliocentric radial velocities, it is most likely that they are
main-sequence stars with masses of 3 to 4\,$M_{\odot}$. We note that this agrees 
with arguments also made by \citet{2007ApJ...664..343K}, \citet{2007MNRAS.379.1293Y} and
 \citet{2007arXiv0709.1471B}
   The eight bound HVSs in
Table \ref{tab:HVS} all have positive radial velocities, and positive $v_{\rm
rf}$.  However this is not true for SDSS J172226.55+594155.9, which is dealt
with in the subsection that follows.

\subsection{A returning star?}

SDSS J172226.55+594155.9 is the only star in our sample that has a
comparably large, but negative, $v_{\rm rf}$. The heliocentric radial velocity
is $-476\,\kms$. Is this a bound HVS observed as it is returning towards the
Galactic Centre?  We perform the same investigation as for the other 
HVSs: that is, we set a grid of proper motions and integrate the trajectories
back in time to find if any initial conditions are reconcilable with a
Galactic Centre origin.  We discover that no proper motion will give SDSS
J172226.55+594155.9 a trajectory that passes through the Galactic Centre.
Given its current radial velocity and position we can find a proper motion such
that the star will pass through the Galactic Centre in the {\it future}.
However such a trajectory is always unbound given a reasonable mass of the galactic halo. 
Making this star bound would require a halo many orders of magnitude more massive than suggested in literature, corresponding to a radius large enough to include other galaxies.
   Given this, SDSS J172226.55+594155.9 can not be a returning, bound HVS.

As SDSS J172226.55+594155.9 is not a HVS, we cannot find its proper motion by
constraining its trajectory to pass through the Galactic Centre.  We consider
the case where its velocity tangential to the line of sight is small compared to
its radial velocity and integrate the trajectory taking its velocity to be equal
to the observed radial velocity.  The trajectory under these assumptions
resembles a typical halo object, extending to 45\,kpc in $R$ and 40\,kpc in $z$.
Hence the observed radial velocity of SDSS J172226.55+594155.9 is consistent
with it being a halo object.
Indeed this can also be the case for some of the stars we claim to be bound. 
However, the velocity distribution observed by \citet{2007ApJ...660..311B} 
suggests that this cannot be the case for all of them.

\subsection{The captured stars}

The suggestion of \citet{2003ApJ...592..935G} that the captured stars are the 
observed S-stars in the galactic centre is compared with our data. As seen from 
Equation~\ref{eq:P_ej}, for encounters between binaries and the
SMBH that eject stars that form HVSs we expect the Hills parameter $D$ to lie
between 0 and 175.  For an ejection probability of 0.5 we have $D=88$.  Taking
this typical value we obtain $f_{\rm r}$ from Equation \ref{eq:f_r} and then the
separation of the binary that produced the HVS from Equation \ref{eq:vej}. The
binary separation and the assumed $D$ allow us to calculate the closest approach
to the black hole, $R_{\rm min}$, via Equation \ref{eq:D}. We find that $R_{\rm
min}$ is typically 50\,AU.

If we further assume that the energy in the binary is small compared to the
potential and kinetic energies post ejection, the binding energy of the captured
star is comparable to the kinetic energy of the ejected star, hence writing the
semi-major axis of the captured star's orbit as $a_{\rm bh}$ we obtain
\begin{equation}
\label{eq:a}
\frac{G M_{\rm bh} M_{\rm s} }{2 a_{\rm bh}} \simeq \frac{M_{\rm hvs}
  v_{\rm ej}^2}{2}.
\end{equation}
With typical masses for the bound HVS and S-star, $M_{\rm hvs}\simeq M_{\rm s}\simeq3 \Mo$ and a typical
ejection velocity of about 800 km/s, we calculate the semimajor axis of the
black hole - captured star system as $a_{\rm bh}\simeq5000AU$. The eccentricity of the
captured star's orbit is given by
\begin{equation}
e=1-\frac{R_{\rm min}}{a_{\rm bh}},
\end{equation}
and is typically very large, $0.99$.  These values are similar to those for some of the
observed S-stars, see \citet{2005ApJ...628..246E} and \citet{2005ApJ...620..744G}. 
A number of the S-stars have eccentricities significantly lower than this, suggesting that
if they were produced by tidal disruption of a binary they were produced in encounters with $R_{\rm min}$ larger than 50\,AU. 

\section{Conclusions}

We assume that the twelve hypervelocity stars with \vrf{} between $275\,{\rm
km\,s^{-1}}$ and $450\,{\rm km\,s^{-1}}$, listed in Table~\ref{tab:HVS}, come
from the Galactic Centre.  Hence we have obtained their proper motions by
constraining the trajectory that takes them to their current position and space
velocity to have originated in the Galactic Centre.

From this present-day space velocity we evaluate whether the
stars are bound to the Galaxy.  The range of proper motions that are consistent
with a Galactic Centre origin are sufficiently small to allow us to
unambiguously determine whether the stars are bound or not.  We find that eight
of our sample of twelve stars are bound to the Galaxy.

We further obtain the velocity with which the stars were ejected from the
Galactic Centre, $v_{\rm ej}$.  These velocities have a range of magnitudes
between $750\,\kms$ and $900\,\kms$.  The unbound stars have similar but
higher kick velocities, consistent with them having the same origin.

We integrate the trajectories of the bound stars forward in time in order to
calculate their return time.  Were the lifetimes of these objects greater or
equal to the return times we would observe an equal number of systems moving
towards the Galactic Centre as away from it.  As we do not see any such
returning objects we conclude that the lifetimes of the observed HVSs must be
much shorter than their return times, and hence that they must be massive
main-sequence stars.

One system, SDSS J172226.55+594155.9, is observed moving towards the Galactic
Centre with a high velocity.  We find that there are no trajectories consistent
with it having been produced in the Galactic Centre.  We calculate its
trajectory if given a small proper motion and conclude that this is consistent
with it being a halo object.

The production rate of bound, high-mass HVSs is about one system per 15\,Myr.
These stars are produced by the breakup of a binary, where the other star is
left bound to the central supermassive black hole.  On the assumption that the
binaries have a mass ratio roughly equal to unity we show that this is
consistent with the other star becoming one of the S-stars.

\section*{Acknowledgements}
The authors would like to thank W.~R.~Brown for generously providing data on
SDSS J172226.55+594155.9.  RPC would like to thank the Swedish Institute for a
Guest Scholarship.  MBD is a Royal Swedish Academy Research Fellow supported by
a grant from the Knut and Alice Wallenberg Foundation.

\bibliographystyle{mn2e}
\bibliography{paper}

\begin{thebibliography}{15}
\expandafter\ifx\csname natexlab\endcsname\relax\def\natexlab#1{#1}\fi

\bibitem[{{Bromley} et~al.(2006){Bromley}, {Kenyon}, {Geller}, {Barcikowski},
  {Brown}, \& {Kurtz}}]{2006ApJ...653.1194B}
{Bromley}, B.~C., {Kenyon}, S.~J., {Geller}, M.~J., {Barcikowski}, E., {Brown},
  W.~R., {Kurtz}, M.~J., 2006, \apj, 653, 1194

\bibitem[{{Brown} et~al.(2006{\natexlab{a}}){Brown}, {Geller}, {Kenyon}, \&
  {Kurtz}}]{2006ApJ...640L..35B}
{Brown}, W.~R., {Geller}, M.~J., {Kenyon}, S.~J., {Kurtz}, M.~J.,
  2006{\natexlab{a}}, \apjl, 640, L35

\bibitem[{{Brown} et~al.(2006{\natexlab{b}}){Brown}, {Geller}, {Kenyon}, \&
  {Kurtz}}]{2006ApJ...647..303B}
{Brown}, W.~R., {Geller}, M.~J., {Kenyon}, S.~J., {Kurtz}, M.~J.,
  2006{\natexlab{b}}, \apj, 647, 303

\bibitem[{{Brown} et~al.(2007{\natexlab{a}}){Brown}, {Geller}, {Kenyon},
  {Kurtz}, \& {Bromley}}]{2007ApJ...660..311B}
{Brown}, W.~R., {Geller}, M.~J., {Kenyon}, S.~J., {Kurtz}, M.~J., {Bromley},
  B.~C., 2007{\natexlab{a}}, \apj, 660, 311

\bibitem[{{Brown} et~al.(2007{\natexlab{b}}){Brown}, {Geller}, {Kenyon},
  {Kurtz}, \& {Bromley}}]{2007arXiv0709.1471B}
{Brown}, W.~R., {Geller}, M.~J., {Kenyon}, S.~J., {Kurtz}, M.~J., {Bromley},
  B.~C., 2007{\natexlab{b}}, ArXiv e-prints, 709

\bibitem[{{Eisenhauer} et~al.(2005)}]{2005ApJ...628..246E}
{Eisenhauer}, F., et~al., 2005, \apj, 628, 246

\bibitem[{{Ghez} et~al.(2005){Ghez}, {Salim}, {Hornstein}, {Tanner}, {Lu},
  {Morris}, {Becklin}, \& {Duch{\^e}ne}}]{2005ApJ...620..744G}
{Ghez}, A.~M., {Salim}, S., {Hornstein}, S.~D., {Tanner}, A., {Lu}, J.~R.,
  {Morris}, M., {Becklin}, E.~E., {Duch{\^e}ne}, G., 2005, \apj, 620, 744

\bibitem[{{Gould} \& {Quillen}(2003)}]{2003ApJ...592..935G}
{Gould}, A., {Quillen}, A.~C., 2003, \apj, 592, 935

\bibitem[{{Hills}(1988)}]{1988Natur.331..687H}
{Hills}, J.~G., 1988, \nat, 331, 687

\bibitem[{{Johnson} \& {Soderblom}(1987)}]{1987AJ.....93..864J}
{Johnson}, D.~R.~H., {Soderblom}, D.~R., 1987, \aj, 93, 864

\bibitem[{{Kollmeier} \& {Gould}(2007)}]{2007ApJ...664..343K}
{Kollmeier}, J.~A., {Gould}, A., 2007, \apj, 664, 343

\bibitem[{{Paczynski}(1990)}]{1990ApJ...348..485P}
{Paczynski}, B., 1990, \apj, 348, 485

\bibitem[{{Wilkinson} \& {Evans}(1999)}]{1999MNRAS.310..645W}
{Wilkinson}, M.~I., {Evans}, N.~W., 1999, \mnras, 310, 645

\bibitem[{{Yu} \& {Madau}(2007)}]{2007MNRAS.379.1293Y}
{Yu}, Q., {Madau}, P., 2007, \mnras, 379, 1293

\bibitem[{{Yu} \& {Tremaine}(2003)}]{2003ApJ...599.1129Y}
{Yu}, Q., {Tremaine}, S., 2003, \apj, 599, 1129

\end{thebibliography}

\label{lastpage}
\end{document}